\documentclass[12pt,preprint]{aastex}
\shorttitle{Collapse Barrier of the Field Clusters}
\shortauthors{Ryu \& Lee}
\begin{document}
\title{Evolution of the Deterministic Collapse Barrier of the Field Clusters as a Probe of Cosmology}
\author{Suho Ryu and Jounghun Lee}
\affil{Astronomy program, Department of Physics and Astronomy,
Seoul National University, Seoul  08826, Republic of Korea \\
\email{jmhera2007@snu.ac.kr, jounghun@astro.snu.ac.kr}}
\begin{abstract}
The collapse barrier, $\delta_{c}$, of the field clusters located in the low-density environment is deterministic rather than diffusive, unlike that of the wall 
counterparts located in the superclusters.  Analyzing the data from the Mira-Titan simulations for eleven different cosmologies including the standard 
$\Lambda$CDM cosmology at various redshifts, we investigate the evolution of the deterministic collapse barrier of the field clusters and explore its dependence 
on the background cosmology. Regardless of the background cosmology, the deterministic $\delta_{c}$ exhibits a universal behavior of having a higher value than 
the Einstein-de Sitter spherical collapse barrier height of $\delta_{sc}=1.686$, at $z=0$ but gradually converging down to $\delta_{sc}$ as the dominance of dark 
energy diminishes with the increment of $z$. A significant difference among different cosmologies, however, is found in its convergence rate as well as in 
the critical redshift $z_{c}$ at which $\delta_{c}=\delta_{sc}$. Showing that the convergence rate and critical redshifts can distinguish even between the degenerate 
cosmologies which yield almost identical linear growth factor and cluster mass functions, we suggest that the evolution of the deterministic collapse 
barrier of the field clusters should be a powerful complementary probe of cosmology. 
\end{abstract}
\section{Introduction}\label{sec:intro}

Ever since \citet{PS74} derived an analytic formula for the cluster mass function based on the excursion set theory, its power and 
usefulness as a cosmological probe has been widely demonstrated and well appreciated in the field of the large scale 
structure \citep[e.g.,][]{fan-etal97,WS98,vik-etal09,bas-etal10,IT12,ben-etal13,planck14}.  
The excursion set theory basically depicts the gravitational growth and collapse of an over-dense region into a bound object as a 
{\it random walk process} confined under a barrier whose height is determined by the underlying dynamics. In the original formulation of \citet{PS74} who 
adopted the spherical dynamics, the height of the collapse barrier has a constant value, $\delta_{sc}$, being independent with the cluster mass. 
Various N-body experiments, however, revealed that the original Press-Schechter mass function failed to match well the numerical results at quantitative 
levels, implying the inadequacy of the spherical dynamics \citep[][and references therein]{BM96}.

In the subsequent works which employed more realistic ellipsoidal dynamics to analytically derive the excursion set mass function, 
the height of the collapse barrier was deemed no longer a constant value but a decreasing function of the cluster mass, $M$, to account for the fact 
that the collapse process deviates further from the spherical dynamics on the lower mass scales \citep[e.g.,][]{BM96,smt01,CL01,ST02}. 
Although better agreements with the numerical results were achieved by employing the mass-dependent ellipsoidal collapse barrier, the purely analytic 
evaluation of the cluster mass function had to be relinquished on the ground that no unique condition for the ellipsoidal collapse exists unlike the case 
of the spherical collapse \citep{BM96,CL01,smt01}.  It was required to empirically determine the functional form of the ellipsoidal collapse barrier height 
by fitting the analytic formula to the numerical results, which in turn inevitably weakened the power of the cluster mass function as a probe of cosmology. 
Besides, the high-resolution N-body simulations revealed that even on the fixed mass scale the collapse barrier height exhibited substantial variations 
with the environments as well as with the cluster identification algorithms \citep[e.g.,][and references therein]{rob-etal09}. 
These numerical findings casted down an excursion set based analytic modeling of the cluster mass function, leading the community to acquiesce in relying 
on mere fitting formulae with multiple adjustable parameters \citep[e.g.,][]{tin-etal08}.

The excursion set modeling of the cluster mass function, however, attracted a revived attention when \citet{MR10a,MR10b} brought up an insightful idea 
that the collapse barrier height should be treated as a stochastic variable rather than a deterministic value. Ascribing the diffusive scatters of the 
collapse barrier height to the incessant disturbing influence from the surrounding on the clusters,  \citet{MR10a} successfully incorporated the concept 
of the stochastic barrier height into the excursion set theory with the help of the path integral method and showed that the accuracy of the generalized 
excursion set mass function with stochastic collapse barrier was considerably improved even though it has only a single parameter, $D_{B}$, which 
measures the degree of the stochasticity of $\delta_{c}$ whose ensemble average coincides with $\delta_{sc}$. 

\citet[hereafter, CA]{CA11a} derived a more accurate mass function by extending the formalism of \citet{MR10a} to the ellipsoidal collapse case where 
the ensemble average, $\langle\delta_{c}\rangle$, does not coincide with $\delta_{sc}$ but drifts away from it, depending on the cluster mass scale.  
As a trade-off of introducing an additional parameter, $\beta$, to quantify the deviation of $\langle\delta_{c}\rangle$ from $\delta_{sc}$, 
\citet{CA11a} won two-fold achievement: matching the numerical results as excellently well as pure fitting formula and simultaneously providing much 
deeper physical understanding about the cluster abundance and its evolution \citep[see also][]{CA11b}. Notwithstanding, the efficacy of the generalized excursion 
set mass function as a cosmological diagnostics was not greatly 
elevated by introducing the concept of a {\it stochastically drifting} collapse barrier due to the obscurity in the choice of the joint probability density functions of 
$\delta_{c}$ expressed in terms of the two parameters, $D_{B}$ and $\beta$ \citep[][and references therein]{ach-etal14}.

It was \citet{lee12} who fathomed out that for the case of the field clusters embedded in the lowest-density environments the collapse barrier height would 
behave deterministically (i.e., $D_{B}=0$) since the degree of the surrounding disturbance as well as ambiguity in the identification of the field clusters would be 
negligibly low in the underdense regions. Defining the field clusters as those which do not belong to superclusters, she modified the CA formalism by setting 
$D_{B}=0$ and confirmed its validity against the N-body results at various redshifts for the case of the currently favored $\Lambda$CDM (cosmological constant 
$\Lambda$ and cold dark matter) model. 
The analysis of \citet{lee12} also found a clear trend that the value of $\beta$ gradually dwindles away to $0$ as the redshift $z$ increases, which 
indicates that at some critical redshift, $z_{c}$, the deterministic collapse barrier height, $\delta_{c}$, for the field clusters will become equal to $\delta_{sc}$. 

This trend may be physically understood by the following logics.  The high-$z$ field clusters correspond to the highest peaks in the linear density field whose 
gravitational collapse proceeds spherically \citep{ber94}. At high redshifts $z>0.7$ where the dark matter (DM) density exceeds that of dark energy (DE), 
the universe is well approximated by the Einstein-de Sitter (EdS) cosmology in which $\delta_{sc}=1.686$. 
We speculate that since the convergence rate of the universe to the EdS model is quite susceptible to the background cosmology, the deterministic 
collapse barrier of the field clusters would evolve differently among different cosmologies. 
The aim of this Paper is to examine if the concept of the deterministic collapse barrier for the field clusters is valid even in $w$CDM (dynamical DE 
with equation of state $w$ + CDM) cosmologies  (Sections \ref{sec:review}-\ref{sec:mf_fit}) and to explore whether or not the evolution of $\beta$, i.e., the deviation 
of the deterministic collapse barrier from the EdS spherical collapse value of $\delta_{sc}=1.686$, can be used as a complementary probe of cosmology 
(Section \ref{sec:bz}). 

\section{Abundance of the Field Clusters in Dark Energy Models}

\subsection{A Brief Review of the Analytic Model}\label{sec:review}

The excursion set modeling of the cluster mass function relates the differential number density of the clusters, $dN/d\ln M$, to the 
{\it multiplicity function}, $f(\sigma)$, as \citep{ree-etal03}
\begin{equation}
\label{eqn;exc}
\frac{d\,N(M, z)}{d\,{\rm ln}\,M} =  \frac{\bar{\rho}}{M}\Bigg{\vert}\frac{d\,{\rm ln}\,\sigma^{-1}}{d\,{\rm ln}\,M}\Bigg{\vert}f[\sigma(M, z)]\:,
\label{eqn:exc}
\end{equation}
where $\bar{\rho}$ is the mean matter density at the present epoch, and $\sigma(M, z)$  is the rms density fluctuation of linear density field smoothed 
on the mass scale $M$ at redshift $z$, and $f(\sigma)$ counts the number of the randomly walking overdensities, $\delta$, that just touch the collapse 
barrier, $\delta_{c}$, when the underlying linear density field has the inverse of the rms fluctuation in the differential range of 
$[\ln\sigma^{-1},\ \ln\sigma^{-1}+d\ln\sigma^{-1}]$. The cosmology dependence of $dN/d\ln M$ stems from the dependence of 
$\sigma(M,z)$ on the linear growth factor, $D(z)$, and linear density power spectrum, $P(k)$ as 
$\sigma^{2}(M,z)\propto D^{2}(z) \int^{\infty}_{0} dk\,k^{2}\,P(k)W^{2}(k, M)$ with the spherical top hat window function, $W(k,M)$. 

Assuming that $\delta_{c}$ is a stochastically drifting variable as in \citet{MR10a,MR10b}, the CA formalism approximates the multiplicity function by 
\begin{eqnarray}
\label{eqn:multi}
f_{\rm ca}(\sigma ; D_{B},\beta) &\approx& f^{(0)}(\sigma ; D_{B}, \beta) + 
f^{(1)}_{\beta=0}(\sigma ; D_{B}) + 
f^{(1)}_{\beta}(\sigma ; D_{B}, \beta) + 
f^{(1)}_{\beta^2}(\sigma ; D_{B}, \beta)\, , \\
\label{eqn:f0}
f^{(0)}(\sigma ; D_{B}, \beta) &=& \frac{\delta_{sc}}{\sigma\sqrt{1+D_{B}}} \sqrt{\frac{2}{\pi}}\,
e^{-\frac{(\delta_{sc}+\beta \sigma^2)^2}{2\sigma^2(1+D_{B})}}\, ,\\
\label{eqn:f1b0}
f^{(1)}_{\beta=0}(\sigma ; D_{B}) &=&-\tilde{\kappa}\frac{\delta_{sc}}{\sigma}
\sqrt{\frac{2a}{\pi}}\left[e^{-\frac{a\delta_{sc}^2}{2\sigma^2}}
-\frac{1}{2}\Gamma\left(0,\frac{a \delta_{sc}^2}{2 \sigma^2}\right)\right]\, , \\
\label{eqn:f1b1}
f^{(1)}_{\beta}(\sigma ; D_{B}, \beta) &=&
-\beta\,a\,\delta_{sc}\left[f^{(1)}_{\beta=0}(\sigma; D_B)+\tilde{\kappa}\,
\textrm{erfc}\left(\frac{\delta_{sc}}{\sigma}\sqrt{\frac{a}{2}}\right)\right]\, , \\
\label{eqn:f1b2}
f^{(1)}_{\beta^2}(\sigma ; D_{B}, \beta) &=&\beta^{2}a^{2}\delta^{2}_{sc}\tilde{\kappa}
\biggl\{\textrm{erfc}\left(\frac{\delta_{sc}}{\sigma}\sqrt{\frac{a}{2}}\right)+\\
&& \frac{\sigma}{a\delta_{sc}}\sqrt{\frac{a}{2\pi}}\biggl[e^{-\frac{a\delta_{sc}^2}
{2\sigma^2}}\left(\frac{1}{2}-\frac{a \delta_{sc}^2}{\sigma^2}\right)+\frac{3}{4}\frac{a\delta_{sc}^2}
{\sigma^2}\Gamma\left(0,\frac{a \delta_{sc}^2}{2 \sigma^2}\right)\biggr]\biggr\}\, ,
\end{eqnarray}
with $a\equiv 1/(1+D_B)$, $\tilde{\kappa} = \kappa a$, $\kappa = 0.475$, upper incomplete gamma function $\Gamma(0, x)$ and 
complementary error function ${\rm erfc}(x)$. The statistical properties of the randomly drifting collapse barrier, $\delta_{c}$, are described by the 
two parameters, $D_{B}$ and $\beta$, in Equations (\ref{eqn:multi})-(\ref{eqn:f1b2}). The former, called the diffusion coefficient, is related to   
the scatters of $\delta_{c}$ from its ensemble average, while the latter, called the drifting average coefficient, measures how much the ensemble 
average of $\delta_{c}$ drifts away from the deterministic height of the spherical collapse barrier $\delta_{sc}$ on a given mass scale 
\citep{CA11a,CA11b}. 

\citet{lee12} suggested that for the case of the field clusters the collapse barrier height should be deterministic 
(i.e., $D_{B}=0$) rather than stochastic since the field clusters would experience the least disturbance from the 
surroundings. Setting $D_{B}=0$ in Equation (\ref{eqn:multi}) and putting it into Equation (\ref{eqn:exc}), she modified 
the CA formalism to evaluate the mass function of the field clusters, $dN_{I}/d\ln M$, as 
\begin{equation}
\label{eqn:exc_f}
\frac{dN_{I}(M, z)}{d\ln M} =  \frac{\bar{\rho}}{M}\Bigg{\vert}\frac{d\,{\rm ln}\,\sigma^{-1}}{d\,{\rm ln}\,M}\Bigg{\vert}
f_{\rm ca}\left[\sigma(M,z) ; D_{B}=0,\beta\right] ,
\end{equation}
which has a single coefficient, $\beta$. Empirically determining the values of $\beta$ at three different redshifts ($z=0,0.5,1$) through numerical 
adjustment process, \citet{lee12} confirmed the validity of Equation (\ref{eqn:exc_f}) for the $\Lambda$CDM case.  
In the following Subsections, we will test this analytic model against the numerical results from $N$-body simulations performed 
for various $w$CDM cosmologies and investigate how $\beta$ evolves in different cosmologies. 

\subsection{Comparison with the Numerical Results}\label{sec:mf_fit}

To investigate if Equation (\ref{eqn:exc_f}) can be validly applied to the case of a $w$CDM cosmology where the DE equation of state, $w$, 
evolves with time, we resort to the Mira-Titan simulation conducted by \citet{MT16} on a periodic box of $(2100\,{\rm Mpc})^3$ with $3200^3$ DM 
particles of individual mass $m_{dm}\sim10^{10}\, M_{\odot}$ for 10 different $w$CDM cosmologies (designated as M001, M002, M003, M004, M005, 
M006, M007, M008, M009, M010) as well as for the $\Lambda$CDM case \citep[see also,][]{hab-etal16, hacc19}.  The initial condition of each cosmology was 
specified by seven parameters, $\{\Omega_{m}, \Omega_{b}, h, \sigma_8, n_s, w_0, w_a\}$, 
under the common assumption of a spatially flat geometry ($\Omega_{de}+\Omega_{m}=1$), no neutrino ($\Omega_{\nu}=0$) and evolution of $w$ given as 
$w=w_{0}+w_{a}z/(1+z)$ \citep{CP01,lin03}.  

\begin{figure}
\begin{center}
\includegraphics[scale=0.7]{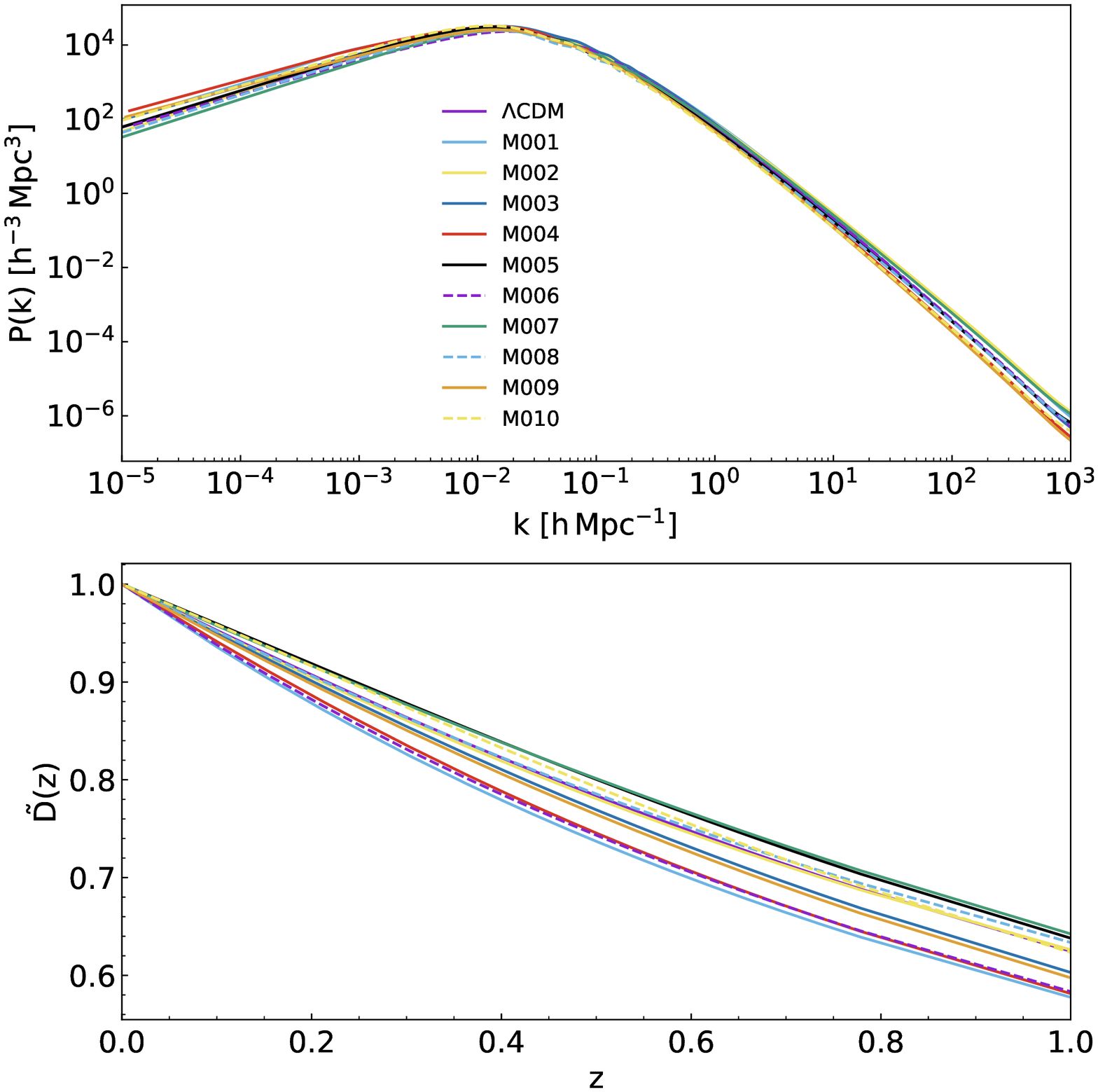}
\caption{Linear density power spectra (top panel) and linear growth factors (bottom panel) from the Mira-Titan simulations for the cases 
of the $\Lambda$CDM and ten different dynamical $w$CDM cosmologies \citep{MT16}.}
\label{fig:pk}
\end{center}
\end{figure}

For the $\Lambda$CDM case ($w_{0}=-1$, $w_{a}=0$), the other five cosmological parameters were set at the best-fit values from the Seven-Year 
Wilkinson Microwave Anisotropy Probe (WMAP7) \citep{wmap7}. For the $w$CDM cosmologies, the values of the seven key cosmological parameters 
including $w_{0}$ and $w_{a}$ were deliberately chosen to be in the ranges that embrace the WMAP7 constraints \citep[for the details, see][]{hei-etal09,MT16}. 
Table \ref{tab:MT_param} lists the values of the key cosmological parameters for each of the eleven different cosmologies from the Mira-Titan 
simulation \citep[see also Table 3 in][]{law-etal17}.  
Figure \ref{fig:pk} plots the linear power spectra at the present epoch, $P(k)$, and the linear growth factor, $D(z)$, for the eleven cosmologies 
(in the top and bottom panels, respectively), computed by the CAMB code \citep{camb}.  Note that the three models, M003, M005 and M008 are almost 
indistinguishable from the $\Lambda$CDM model in $P(k)$, while the two models, M007 and M009, yield $D(z)$ the shapes of which are very similar to that 
for the $\Lambda$CDM case. 

\citet{MT16}  compiled the catalogs of the DM halos identified by applying the friends-of-friends (FoF) algorithm with a linking length of 
$b_{c}\bar{d}_{p}$ with $b_{c}=0.168$ and mean particle separation $\bar{d}_{p}$ to each particle snapshot in the redshift range of $0.0\le z\le 4.0$.
Following the same procedure of \citet{lee12}, we analyze the FoF halo catalogs from each Mira-Titan universe to numerically determine the mass functions 
of the field clusters and the associated errors as well:
\begin{enumerate}
\item
Make a sample of the cluster halos with masses larger than $M_{c}=3\times 10^{13}\,h^{-1}\,M_{\odot}$ out of the halo catalog at a given redshift in the 
range of $0\le z\le z_{c}\sim 1$. The catalogs at higher redshifts, $z>z_{c}$ are excluded from the analysis on the ground that the field clusters at $z>z_{c}$ 
are too rare to yield statistically significant results. 
\item
Apply to the above sample the FoF algorithm with a linking length of $2b_{c}\bar{d}_{c}$ with mean cluster halo separation $\bar{d}_{c}$ to 
find a supercluster as a cluster of clusters each of which consists of two and more cluster halos. This specific choice of the linking length was 
made by \citet{lee12} to guarantee that the degree of the disturbance from the surroundings on the field clusters is indeed negligible (i.e., $D_{B}=0$) 
\citep[see Figure 2 in][]{lee12}.  
\item
Find the cluster halos in the sample which appertain to none of the identified superclusters as the field clusters and count them, $dN_{\rm I}$, 
in the logarithmic mass bin, $[\ln M, \ln M+d\ln M]$.  
\item
Split the field clusters into eight Jackknife subsamples according to their positions and separately determine $dN_{\rm I}/d\ln M$ from each subsample. 
Evaluate the Jackknife errors in the measurement of $dN_{\rm I}/d\ln M$ as one standard deviation scatter around the ensemble 
average over the eight subsamples. 
\end{enumerate}

Now that the mass functions of the field clusters from the Mira-Titan simulations are all determined, we compare them with Equation (\ref{eqn:exc_f}) 
by adjusting the single coefficient, $\beta$. For this comparison, the spherical barrier height, $\delta_{sc}$, is set at the EdS value of $1.686$, since 
it varies only very weakly with the back ground cosmology \citep[e.g.,][]{eke-etal96, pac10}.  We employ the $\chi^{2}$-statistics to determine the best-fit value of 
$\beta$ and estimate the associated error, $\sigma_{\beta}$, as $1/\sqrt{I_{\beta}}$, where $I_{\beta}$ is the Fisher information given as 
$I_{\beta}\equiv d^{2}\chi^{2}/d\beta^{2}$ at the best-fit value of $\beta$, at each redshift for each cosmology.

Figure \ref{fig:mf_z0} plots the numerical result (filled circles) as well as Equation (\ref{eqn:exc_f}) with the best-fit value of $\beta$ (red solid line) 
for eleven different cosmologies at $z=0$. In each panel, the analytic mass function with the best-fit $\beta$ for the $\Lambda$CDM case is shown as dotted line 
for comparison. 
Figures \ref{fig:mf_z0.4}-\ref{fig:mf_z0.8} plot the same as Figure \ref{fig:mf_z0} but at $z=0.4$ and $z=0.78$, respectively. As can be seen, 
Equation (\ref{eqn:exc_f}) with the best-fit $\beta$ is quite successful in matching the numerically determined mass functions of the field clusters for all of 
the eleven cosmologies at all of the three redshifts. As emphasized in \citet{lee12}, the modified CA formalism with $D_{B}=0$ describes well not only the shape 
but also amplitude of the mass function of the field clusters even though it has only a single parameter, $\beta$. 
The good agreements between the analytical and numerical results shown in Figures \ref{fig:mf_z0}-\ref{fig:mf_z0.8} prove that the modified CA formalism 
with the deterministic collapse barrier for the field clusters can be legitimately extended to the $w$CDM cosmologies.

It is, however, worth mentioning here that the analytic model for the field cluster mass function, Equation (\ref{eqn:exc_f}), is found to be valid in the limited 
redshift range $z\le z_{c}$, which we suspect is due to the failure of the assumption $D_{B}=0$ at higher redshifts $z>z_{c}\sim 1$.  The low abundance of the 
clusters with $M\ge M_{c}$ at $z>z_{c}$ makes it difficult to properly identify the superclusters via the FoF algorithm, which in turn contaminates the identification 
of the field clusters. In other words, the field clusters identified via the FoF algorithm at $z>z_{c}$ may not be isolated enough to satisfy the condition of $D_{B}=0$.  
 
\subsection{Evolution of the Drifting Collapse Barrier}\label{sec:bz}

Figure \ref{fig:beta_z} plots the best-fit value of $\beta$ determined in Section \ref{sec:mf_fit} versus $z$ for the eleven cosmologies, revealing the presence 
of a strong anti-correlation between $\beta$ and $z$.  We discover an universal behavior of $\beta(z)$ from all of the eleven cosmologies:   
it monotonically declines toward $0$ as the redshift increases up to $z\ge 1$.  In the range of $0\le z\le 0.3$, it declines relatively slowly 
with $z$, while in the higher $z$-range it drops quite rapidly down to zero. The drifting coefficient, $\beta(z)$, from each of the eleven cosmologies is, 
however, manifestly different from one another in its declining rate and amplitude as well as in the critical redshift at which $\beta(z)$ becomes zero. 

Although $\delta_{sc}/\sigma(M,z)$ may play a partial role to induce the cosmology dependence of $\beta(z)$, we believe that it should not be the main 
contribution. First of all, the spherical collapse barrier height, $\delta_{sc}$, has been known to be quite insensitive to the background cosmology 
as mentioned in Section \ref{sec:mf_fit}. For the case of flat $\Lambda$CDM models, \citet{eke-etal96} showed that $\delta_{sc}$ changes very mildly 
from $1.686$ to $1.67$ as $\Omega_{m}$ changes from $1$ to $0.1$. Even for the case of flat $w$CDM models, the weak dependence of 
$\delta_{sc}$ was rigorously proven by Pace et al. (2010) who directly solved the nonlinear differential equation of the density contrast in the 
spherical collapse process to find that the value of $\delta_{sc}(z)$ for the $w$CDM models remain very similar to that for the $\Lambda$CDM 
model in the whole redshift range. 

Regarding the cosmology dependence of $\sigma(M,z)$,  it depends on the background cosmology only through $D(z)$ and $P(k)$. Whereas, as can be seen 
in Figure \ref{fig:beta_z}, $\beta(z)$ differs even among those models which have the same shapes of $D(z)$ and $P(k)$. 
Therefore, the cosmology dependence of $\beta(z)$ witnessed in Figure \ref{fig:beta_z} should come mainly from another channel, which we believe is the departure of 
$\delta_{c}$ from $\delta_{sc}$. 
In different cosmologies, the non-spherical collapse in the nonlinear regime would proceed differently, resulting in the cosmology dependence of the degree 
of the departure of $\delta_{c}$ from $\delta_{sc}$, which is described by the single parameter, $\beta(z)$, for the case of the field cluster abundance. 

Without having a physical model for the effect of the background cosmology on the departure of $\delta_{c}$ from $\delta_{sc}$ at the moment, we find the following fitting 
formula useful to quantitatively describe the ways in which $\beta(z)$ differs among the eleven cosmologies and to efficiently assess the statistical significances  
of their differences: 
\begin{equation}
\beta(z) = \beta_{A}\ {\sinh}^{-1} \left[\frac{1}{q_z}(z-z_c)\right],
\label{eqn:inv_sinh} 
\end{equation}
where three adjustable parameters, $\beta_{A}$, $q_{z}$ and $z_{c}$, denote the amplitude, redshift dispersion and critical redshift of $\beta(z)$, respectively. 
The overall amplitude, $\beta_{A}$,  quantifies how much $\delta_{c}$ departs from the EdS value of $\delta_{sc}$ at $z=0$, the critical redshift parameter, $z_{c}$, 
quantifies when $\delta_{c}$ becomes equal to $\delta_{sc}$, while the inverse of the redshift dispersion, $1/q_{z}$, quantifies the rate at which $\delta_{c}$ 
converges to  $\delta_{sc}$, as $z$ increases.  
The best-fit values of $(\beta_{A}, q_{z}, z_{c})$ and their associated errors $(\sigma_{\beta_{A}},\sigma_{q_{z}}, \sigma_{z_c})$ are obtained by fitting 
Equation (\ref{eqn:inv_sinh}) to the empirically determined $\beta(z)$ in Section \ref{sec:mf_fit} with the help of the ordinary least square code 
(see Table \ref{tab:beta_zc}). 

Figure \ref{fig:beta_fit} shows how well Equation (\ref{eqn:inv_sinh}) with three best-fit parameters (red solid line) describes the empirically determined 
$\beta(z)$ (filled circles), comparing the best-fit $\beta(z)$ for each of the ten $w$CDM cosmologies with that for the $\Lambda$CDM case (dotted line).  
It is interesting to see that the three cosmologies, $\Lambda$CDM, M007, and M009, which produce almost identical mass functions of the field clusters at all 
redshifts (Figures \ref{fig:mf_z0}-\ref{fig:mf_z0.8}), can still be distinguished by their distinct $\beta(z)$. The differences in the best-fit values of the critical 
redshifts, $\Delta_{zc}$, between the $\Lambda$CDM and M007 (M009) cases is as high as $3.47\sigma_{\Delta_{zc}}$ ($5.89\sigma_{\Delta_{zc}}$). 
Here, the errors, $\sigma_{\Delta_{zc}}$ is calculated through the error propagation as 
$\sigma_{\Delta_{zc}}\equiv \left(\sigma^{2}_{zc,1}+\sigma^{2}_{zc,2}\right)^{1/2}$ where $\sigma_{zc,1}$ and $\sigma_{zc,2}$  
are the errors in the measurements of $z_{c}$ for the $\Lambda$CDM and M007 (M009) cases, respectively. 
Note also that $\beta(z)$ can also distinguish between the two cosmologies, M002 and $\Lambda$CDM, although both of the cosmologies yield quite 
similar linear growth factors and field cluster mass functions (Figure \ref{fig:pk}). 
The difference, $\Delta_{zc}$, between the two cosmologies is found to be as significant as $14\sigma_{\Delta_{zc}}$.

The evolution of $\beta(z)$ also allows us to distinguish not only between the $w$CDM and $\Lambda$CDM cosmologies but also among different 
$w$CDM cosmologies themselves.  For instance,  the two $w$CDM cosmologies, M001 and M006, are found to have almost no difference in 
their field cluster mass functions. Nevertheless, they can be distinguished by the $6.7\sigma_{\Delta_{zc}}$ differences in the best-fit values of $z_{c}$. 
These results clearly indicates a potential of $\beta(z)$ to complement the cluster mass function in discriminating the candidate cosmologies.

\section{Summary and Discussion}

\begin{figure}
\begin{center}
\includegraphics[scale=0.7]{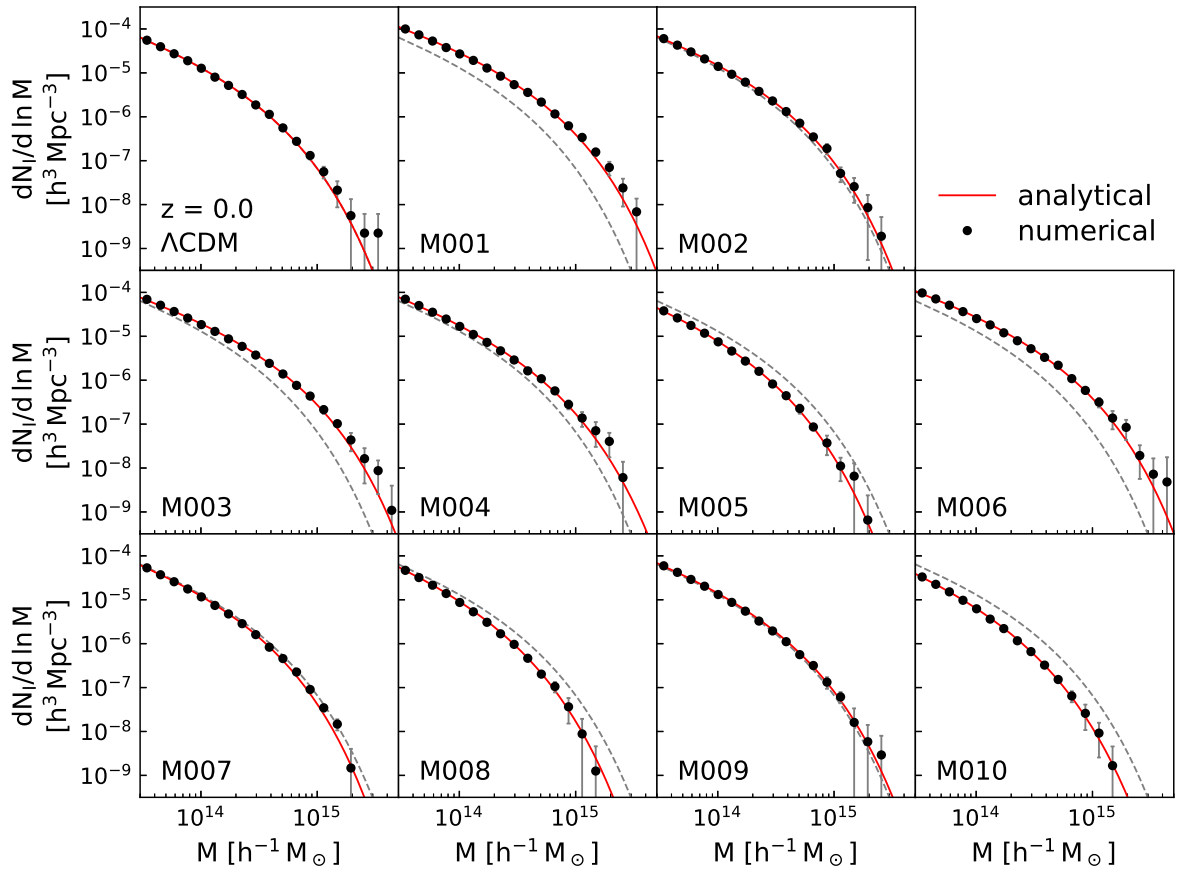}
\caption{Numerically obtained mass functions of the field clusters (filled circles) compared with the 
analytic formula (red solid lines) for $10$ different dynamical $w$CDM cosmologies as well as for the $\Lambda$CDM case at $z=0$.}
\label{fig:mf_z0}
\end{center}
\end{figure}

\begin{figure}
\begin{center}
\includegraphics[scale=0.7]{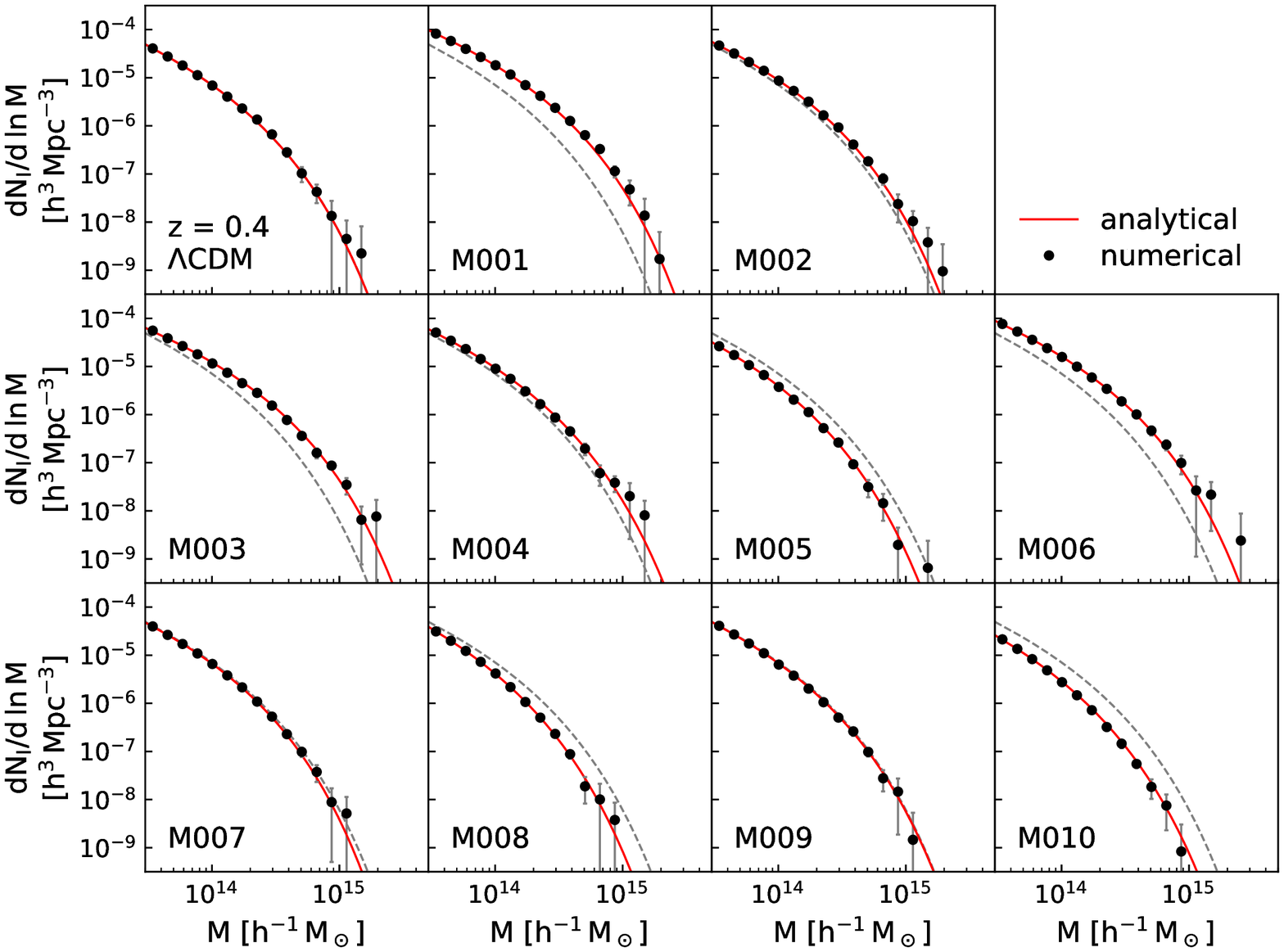}
\caption{Same as Figure \ref{fig:mf_z0} but for at $z=0.4$.}
\label{fig:mf_z0.4}
\end{center}
\end{figure}

\begin{figure}
\begin{center}
\includegraphics[scale=0.7]{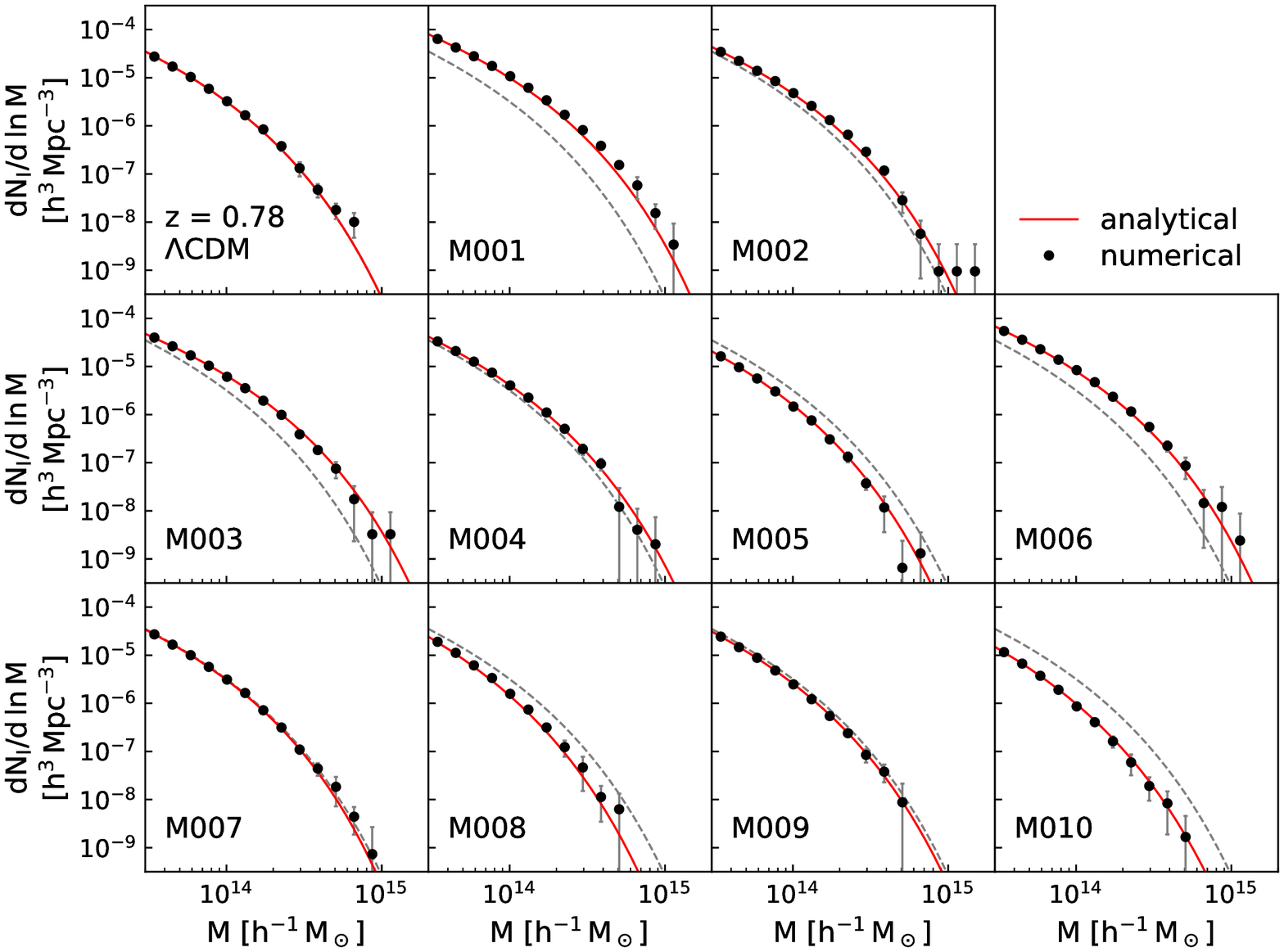}
\caption{Same as Figure \ref{fig:mf_z0} but for at $z=0.78$.}
\label{fig:mf_z0.8}
\end{center}
\end{figure}

\begin{figure}
\begin{center}
\includegraphics[scale=0.7]{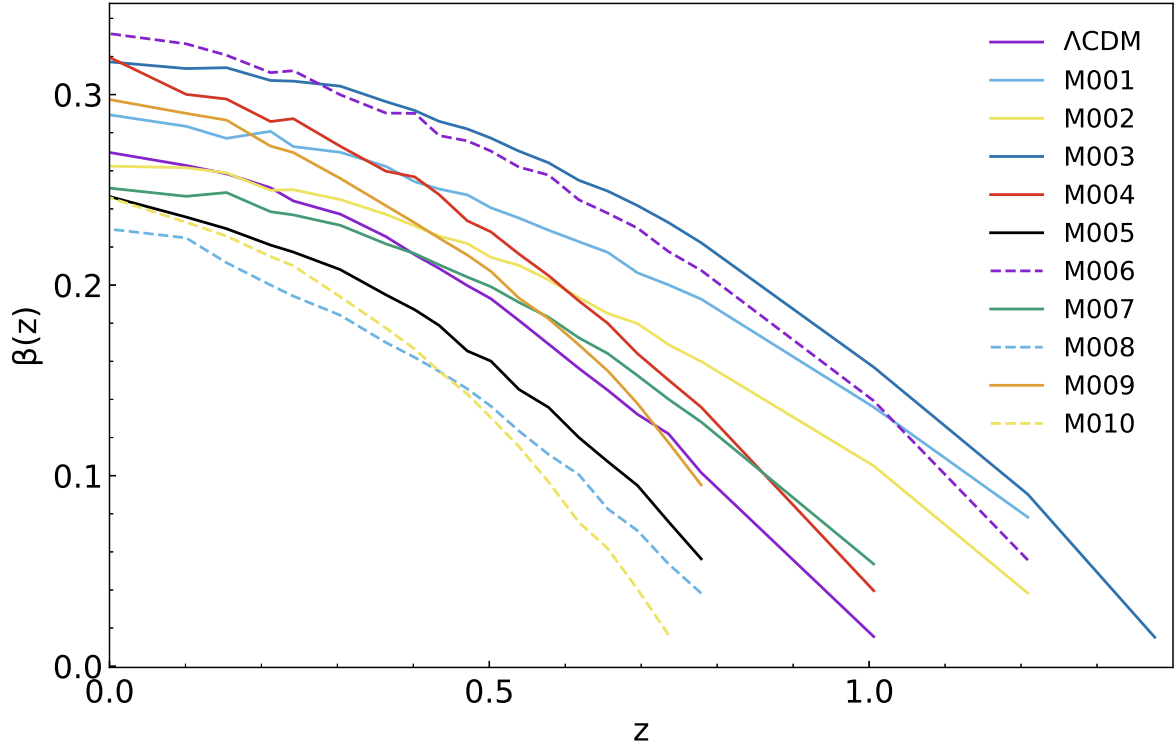}
\caption{Redshift evolution of the drifting coefficient, $\beta$, for $11$ different DE cosmologies.}
\label{fig:beta_z}
\end{center}
\end{figure}

Numerically determining the field cluster mass functions at various redshifts from the Mira-Titan simulations for eleven different DE cosmologies 
(ten different $w$CDM and one $\Lambda$CDM cosmologies) whose key cosmological parameters are chosen to be in the range covering well the WMAP7 
constraints, we have shown that the numerical results at all redshifts for all eleven cosmologies agree very well with the analytic model obtained 
by \citet{lee12} through a modification of the generalized excursion set formalism (Figure \ref{fig:mf_z0}-\ref{fig:mf_z0.8}).  
The success of the analytic model has validated the key assumptions of \citet{lee12} that for the field clusters the collapse barrier 
can be deemed deterministic and thus that their excursion set mass function can be fully characterized by a single {\it drifting coefficient}, $\beta$, which 
measures the degree of the departure of the collapse barrier height from the spherical height, $\delta_{sc}$. 
It has been found that $\beta(z)$ exhibits a universal tendency of converging to zero with the increment of $z$ and that its convergence rate 
as well as the value of critical redshift, $z_{c}$ at which $\beta(z)=0$ depends strongly on the background cosmology (Figures \ref{fig:beta_z}). 
Noting that $\beta(z)$ differs even among those cosmologies that are degenerate with one another in the linear power spectrum, linear growth 
factor and cluster mass function, we suggest that $\beta(z)$ should be in principle useful to discriminate the candidate cosmologies. 

Nevertheless, since the eleven Mira-Titan cosmologies differ not only in their DE equation of states ($w_{0},w_{a}$) and DE density parameters ($\Omega_{de}$) 
but also in the values of the other five key cosmological parameters ($h,\Omega_{m},\Omega_{b},n_{s},\sigma_{8}$), the detected strong cosmology 
dependence of $\beta(z)$ cannot be entirely ascribed to the differences among the eleven models in the values of $w_{0},w_{a}$ and $\Omega_{de}$. 
In other words, our work has demonstrated the usefulness of $\beta(z)$ as a discriminator of $w$CDM cosmologies from the $\Lambda$CDM 
model, but not as a complementary probe of DE equation of state.   

\begin{figure}
\begin{center}
\includegraphics[scale=0.7]{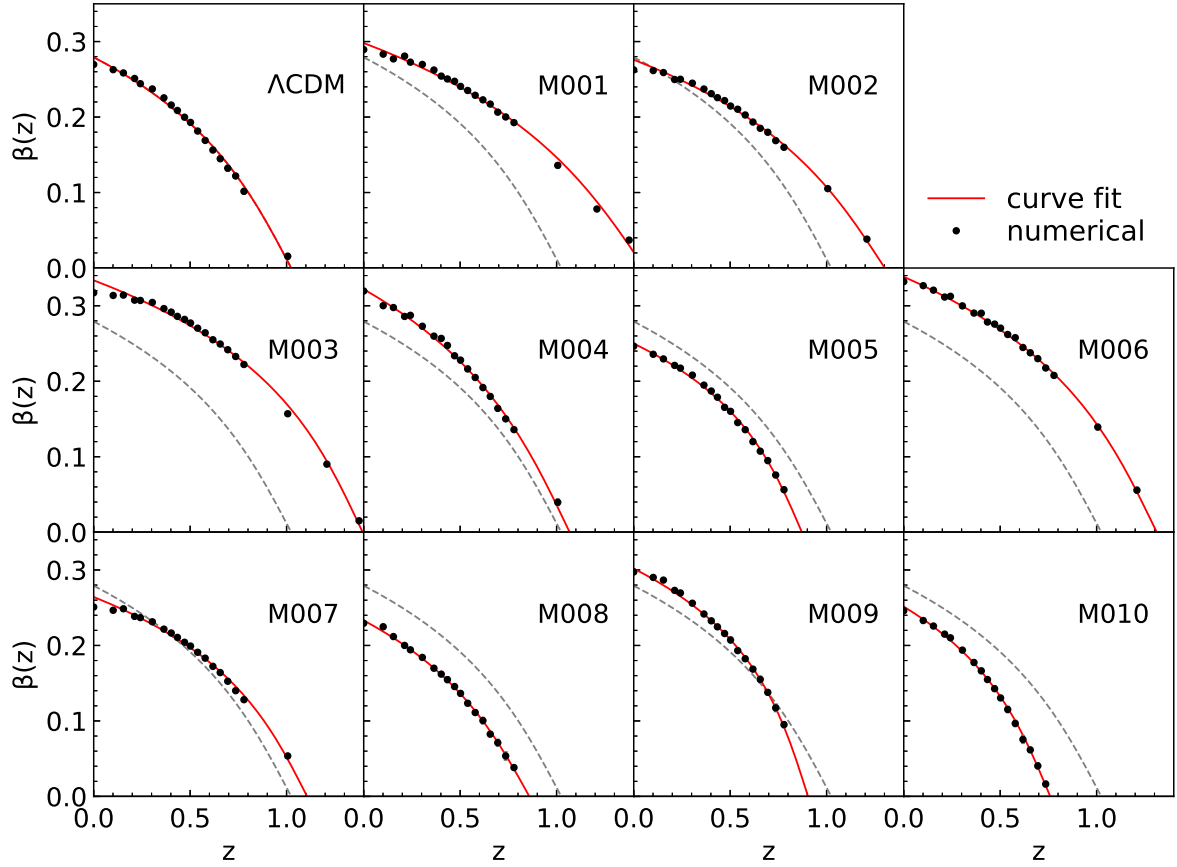}
\caption{Linear fits (red solid lines) to the numerically obtained $\beta(z)$ (filled circles) for $11$ different DE cosmologies.}
\label{fig:beta_fit}
\end{center}
\end{figure}

A more comprehensive investigation should be carried out to sort out the sole effect of the DE equation of state on $\beta(z)$ before claiming it as a probe 
of DE in practice. What will be highly desirable is to examine how sensitively $\beta(z)$ reacts to the variations of the DE equation of state and density parameter by 
determining its shapes from a series of N-body simulations each of which has a different DE equation of state but the same values of the other key cosmological 
parameters.
What will be even more highly desirable is to construct a theoretical formula for $\beta(z)$ from a physical principle. Although Equation (\ref{eqn:inv_sinh}) is a 
mere fitting formula expressed in terms of an inverse sine hyperbolic function with three adjustable parameters, its general success in matching $\beta(z)$ 
for all of the eleven cosmologies (Figure \ref{fig:beta_fit}) hints a prospect for finding a physical formula similar to it and directly linking its three parameters to the 
initial conditions.  This physical formula, if found and verified to be robust, would allow us to probe not only the DE equation of state and density parameter but also 
the other alternative cosmologies such massive neutrinos, modified gravity and etc, with $\beta(z)$. Our future work is in this direction. 

\acknowledgments
We thank the anonymous referee for very useful comments which helped us improve the original manuscript.
We acknowledge the support by Basic Science Research Program through the National Research Foundation (NRF) of Korea 
funded by the Ministry of Education (No.2019R1A2C1083855) and also by a research grant from the NRF to the Center for Galaxy 
Evolution Research (No.2017R1A5A1070354). 

An award of computer time was provided by the Innovative and Novel Computational Impact on Theory and Experiment (INCITE) program. This research used 
resources of the Argonne Leadership Computing Facility, which is a DOE Office of Science User Facility supported under contract DE-AC02-06CH11357. This 
research also used resources of the Oak Ridge Leadership Computing Facility, which is a DOE Office of Science User Facility supported under Contract DE-
AC05-00OR22725.

\clearpage

\clearpage
\clearpage
\clearpage
\clearpage
\begin{deluxetable}{cccccccc}
\tablewidth{0pt}
\setlength{\tabcolsep}{3mm}
\tablecaption{Key cosmological parameters for the eleven models from the HACC simulations}
\tablehead{Cosmology & $\Omega_{m}$ & $\Omega_{b}$ & $h$ & $\sigma_8$ & $n_s$ & $w_0$ & $w_a$}
\startdata
${\rm \Lambda CDM}$ & 0.2648 & 0.04479 & 0.7100 & 0.8000 & 0.9630 & -1.0000 & 0.0000\\
M001 & 0.3871 & 0.05945 & 0.6167 & 0.8778 & 0.9611 & -0.7000 & 0.6722\\
M002 & 0.2411 & 0.04139 & 0.7500 & 0.8556 & 1.0500 & -1.0330 & 0.9111\\
M003 & 0.3017 & 0.04271 & 0.7167 & 0.9000 & 0.8944 & -1.1000 & -0.2833\\
M004 & 0.3642 & 0.06710 & 0.5833 & 0.7889 & 0.8722 & -1.1670 & 1.1500\\
M005 & 0.1983 & 0.03253 & 0.8500 & 0.7667 & 0.9833 & -1.2330 & -0.0445\\
M006 & 0.4354 & 0.07107 & 0.5500 & 0.8333 & 0.9167 & -0.7667 & 0.1944\\
M007 & 0.2265 & 0.03324 & 0.8167 & 0.8111 & 1.0280 & -0.8333 & -1.0000\\
M008 & 0.2570 & 0.04939 & 0.6833 & 0.7000 & 1.0060 & -0.9000 & 0.4333\\
M009 & 0.3299 & 0.05141 & 0.6500 & 0.7444 & 0.8500 & -0.9667 & -0.7611\\
M010 & 0.2083 & 0.03649 & 0.7833 & 0.7222 & 0.9389 & -1.3000 & -0.5222
\enddata
\label{tab:MT_param}
\end{deluxetable}
\clearpage
\begin{deluxetable}{cccc}
\tablewidth{0pt}
\setlength{\tabcolsep}{5mm}
\tablecaption{Best-fit parameters for the evolution of the drifting coefficient.}
\tablehead{Cosmology & $\beta_{A}$ & $q_z$ & $z_c$}
\startdata
${\rm \Lambda CDM}$  	& $-0.141\pm 0.008$ 	& $0.289\pm 0.033$ 	& $1.024\pm 0.014$ \\
${\rm M001}$  	& $-0.147\pm 0.008$ 	& $0.388\pm 0.045$ 	& $1.456\pm 0.018$ \\
${\rm M002}$  	& $-0.135\pm 0.005$ 	& $0.343\pm 0.026$ 	& $1.302\pm 0.014$ \\
${\rm M003}$  	& $-0.138\pm 0.006$ 	& $0.252\pm 0.026$ 	& $1.394\pm 0.011$ \\
${\rm M004}$  	& $-0.163\pm 0.008$ 	& $0.303\pm 0.032$ 	& $1.068\pm 0.014$ \\
${\rm M005}$  	& $-0.111\pm 0.005$ 	& $0.186\pm 0.018$ 	& $0.872\pm 0.009$ \\
${\rm M006}$  	& $-0.152\pm 0.005$ 	& $0.285\pm 0.021$ 	& $1.311\pm 0.012$ \\
${\rm M007}$  	& $-0.116\pm 0.007$ 	& $0.229\pm 0.031$ 	& $1.106\pm 0.019$ \\
${\rm M008}$  	& $-0.124\pm 0.005$ 	& $0.269\pm 0.021$ 	& $0.859\pm 0.007$ \\
${\rm M009}$  	& $-0.120\pm 0.006$ 	& $0.147\pm 0.019$ 	& $0.903\pm 0.015$ \\
${\rm M010}$  	& $-0.123\pm 0.004$ 	& $0.199\pm 0.013$ 	& $0.759\pm 0.005$ 

\enddata
\label{tab:beta_zc}
\end{deluxetable}
\end{document}